# Translating XPS Measurement Procedure for Band Alignment into Reliable *Ab-initio* Calculation Method

Daoyu Zhang,*,† Minnan Yang,‡ Huimin Gao,† and Shuai Dong*,†

†School of Physics, Southeast University, Nanjing, 211189, China.

‡Department of Physics, China Pharmaceutical University, Nanjing, 211198, China.

**ABSTRACT:** Band alignment between solids is a crucial issue in condensed matter physics and electronic devices. Although the XPS method has been used as a routine method for determination of the band alignment, the theoretical calculations by copying the XPS band alignment procedure usually fail to match the measured results. In this work, a reliable *ab-initio* calculation method for band alignment is proposed on the basis of the XPS procedure and in consideration of surface polarity and lattice deformation. Application of our method to anatase and rutile $TiO_2$ shows well agreement between calculation and experiment. Furthermore, our method can produce two types of band alignment: the coupled and the intrinsic, depending on whether the solid/solid interface effect is involved or not. The coupled and intrinsic band alignments correspond to alignments measured by XPS and electrochemical impedance analysis, respectively, explaining why band alignments reported by these two experiments are rather inconsistent.

The energy band alignment between solids is a fundamental concept in condensed matter physics and of crucial importance in design of electronic devices.[1-4] Many experimental methods, such as X-ray photoelectron spectroscopy (XPS), electrochemical impedance analysis (EIA), photoluminescence spectroscopy, surface photovoltage measurements, scanning probe microscopy, and so on, have been developed to measure the band alignment, the band offset, as well as the band bending. Among these methods, XPS is one of the most reliable and widely adopted choice.[5] Using XPS, the valence band offset between solids A and B, $\Delta E_{B-A}(VBM)$, can be expressed as:[6]

$$\Delta E_{B-A}(VBM) = [\Delta E_{VBM-CL}(B) - \Delta E_{VBM-CL}(A)] + \Delta E_{B-A}(CL) \quad (1)$$

in which VBM stands for the valence band maximum and CL for the core level; $\Delta E_{Y-X}(Z)$ here and hereinafter denotes the energy difference of level Z between systems Y and X, or the energy difference between levels Y and X in the system Z.

Unfortunately, directly employing Eq. 1 to calculate the band-offset using density functional theory (DFT) usually does not correctly reproduce the experimental values. For example, Wei *et al.* calculated the band offsets between all II-VI and III-V semiconductor compounds copying the XPS measurement procedure and found that some band offsets have large deviations from the experimental observations, especially between compounds with unavoidable lattice mismatch.[7] Chambers *et al.* studied the band offset at the epitaxial anatase $TiO_2/n$-$SrTiO_3$(001) interface, and their XPS measurement and DFT calculation showed bad agreement between theory and experiment.[8] Thus they concluded that either DFT could not accurately calculate band offsets in the oxide materials, or some unknown factors (missed in the modeling) at the interface were influencing the band offset.

In fact, DFT is presently the most successful approach to compute the electronic structure of matter. Although the local density approximation (LDA) or generalized gradient approximation (GGA) will underestimate the correlation effects between electrons and the band gap, it remains anticipated to calculate band alignments between solids. Since the band alignment between solids is a relative energy value, the LDA or GGA error can be largely canceled in calculations.[9] For example, Ju *et al.* obtained the same band offsets between anatase and rutile $TiO_2$ by using the functional GGA and a more accurate hybrid exchange-correlation (XC) functional HSE06.[10] Therefore, the aforementioned failure of DFT prediction of band alignment should be due to other reasons.

In this work, we attempt to translate the XPS measurement procedure for energy band alignment between solids to a reliable *ab-initio* calculation method. The effects from lattice mismatch and surface polarity are taken into account. The calculation procedure for the band alignment between solids A and B is illustrated in Fig. 1. $\Delta E_{VBM-CL}(A)$ and $\Delta E_{VBM-CL}(B)$ terms in Eq. 1 can be calculated at their respective equilibrium lattice constants. The $\Delta E_{B-A}(CL)$ term can be obtained from the unrelaxed A'/B' heterojunction, but two corrections are needed for this term.

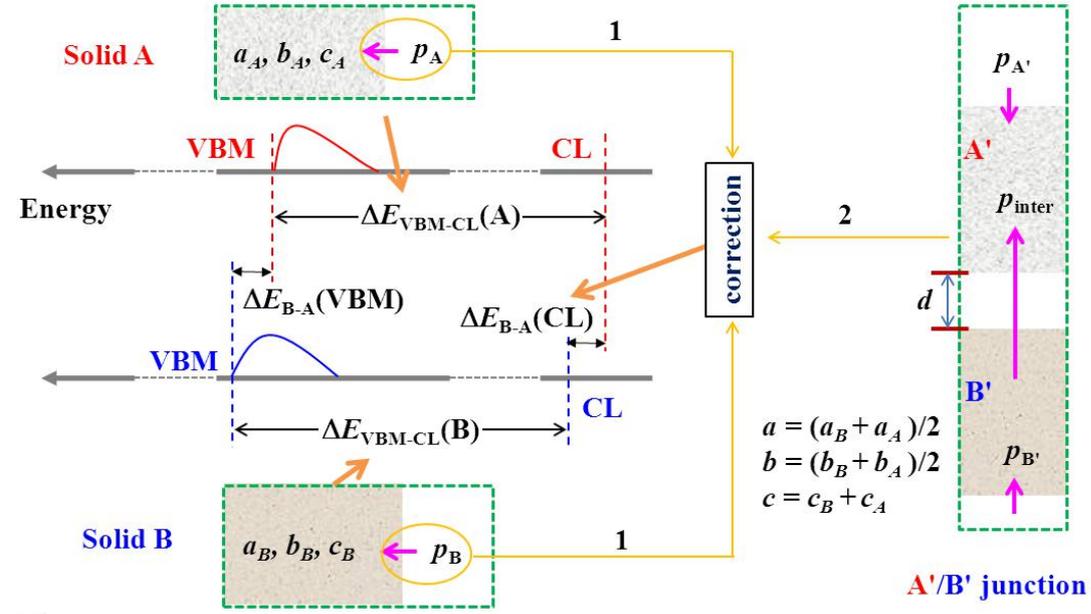

**Figure 1.** Schematic diagram of the procedure for calculating the band alignment between solids A and B, based on the XPS measurement procedure (Eq. 1). $\Delta E_{VBM-CL}(A)$ and $\Delta E_{VBM-CL}(B)$ can be calculated at their respective equilibrium lattice constants, while $\Delta E_{B-A}(CL)$ can be obtained from the A'/B' junction with corrections 1 and 2 (from the surface polarity and the expansion and/or compression of A and B respectively). The surface polarity, which is relative to the bulk one, is described by the electric dipole moment $p$. $p_{inter}$ denotes the moment of the dipole layer induced by the interaction between two solids. The distance between two solids $d$ can be tuned to change their interaction. When $d$ is big enough, $p_{inter}$ equals to zero, and the band alignment between two solids is intrinsic; the nonzero $p_{inter}$ will give a coupled band alignment.

The first correction is from the surface polarity, which is introduced by the interruption of lattice periodicity. The polarity of an exposed surface may be represented by electric dipoles (moment $p$) perpendicular to the surface, as shown in Fig. 2(a). Under the parallel-plate capacitor approximation the electrostatic potential difference ($\varphi$) between two sides of the capacitor can be expressed as:

$$\varphi = ep/(A\varepsilon_{eff}\varepsilon_0), \qquad (2)$$

where $e$ is the elementary charge; $A$ is the surface area; $\varepsilon_{eff}$ is the effective dielectric constant of the surface; and $\varepsilon_0$ is the permittivity of free space. If the average electrostatic potential of the system is taken as the energy reference, then the electrostatic potential created by the surface dipole layer increases the potential of $\Delta E^{VR}$ in the vacuum side of the capacitor and decreases the potential of $\Delta E^{SR}$ in the solid side. Fig. 2(b) shows the superposition of this additional electrostatic potential from the surface dipole layer and the original electrostatic potential of the solid. The solid line in Fig. 2(b) is the resulting electrostatic potential of the solid with an exposed surface. Thus, the surface polarity will lead to an increment of $\varphi$ ($\varphi = \Delta E^{VR} - \Delta E^{SR}$) in the binding energy of every electron in the solid. In other words, the energy levels of both valence and core electrons in the solid will be shifted by the same amount of $-\varphi$ due to the surface polarity. The effect of surficial polarity on the energy levels might as well be put into the $\Delta E_{B-A}(CL)$ term in Eq. 1, and the first correction is:

$$\Delta E_{B-A}(CL)_1 = [-\varphi(B)] - [-\varphi(A)]. \tag{3}$$

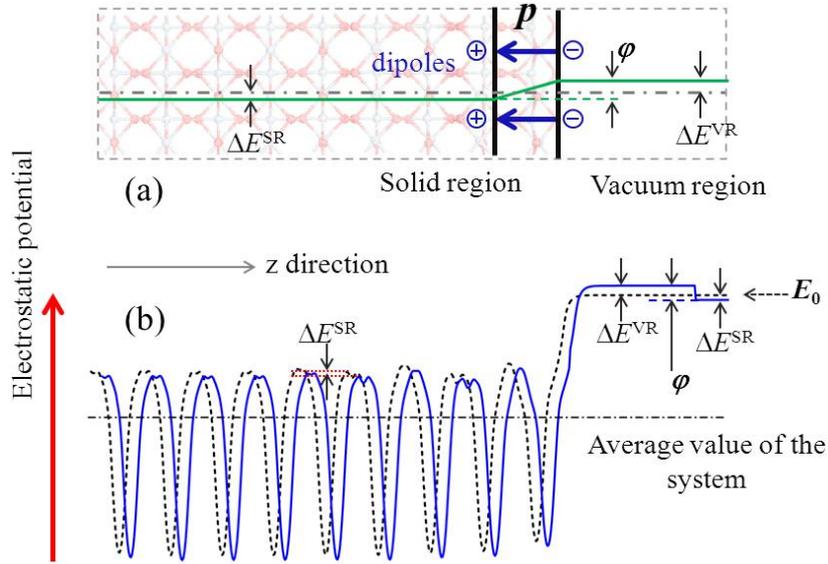

**Figure 2**. (a) The polarity surface of a solid and its polarity is represented by the electric dipole layer with the dipole moment of magnitude $p$. Under the parallel-plate capacitor approximation the electrostatic potential difference between the two plates is $\varphi$. If the average electrostatic potential of the system is taken as the energy reference, then the potential in the solid region decreases of $\Delta E^{SR}$ and that in the vacuum region increases of $\Delta E^{VR}$. (b) Superposing the electrostatic potential of the surface dipole layer on the electrostatic potential of the solid, which is averaged in the $xy$ plane parallel to the surface[11]; the dashed line indicates the potential without the surface dipoles.

The second correction comes from the lattice mismatch between solids when constructing the heterojunction. In theoretical calculation, the in-plane lattice should be expanded or compressed from original solids A and B to new A' and B' to form a junction supercell (see Fig. 1). Thus the core level difference between two solids calculated from the A'/B' junction is $\Delta E_{B'-A'}(CL)$, but evidently not the original $\Delta E_{\text{B-A}}(\text{CL})$.

The lattice deformation (i.e. the lattice expansion and/or compression) can lead to a change in the local atomic environment, and thus causes shifts of the core levels. For a given solid X and its deformed form X', their core levels $cl(\text{X})$ and $cl(\text{X}')$ can be obtained from their respective *ab initio* calculations. Then the change of the core level from the lattice deformation can be determined by:

$$\Delta E_{\text{X'-X}}(\text{CL}) = [cl(\text{X}') + \varepsilon_{\text{Fermi}}(\text{X}') - E_0(\text{X}')] - [cl(\text{X}) + \varepsilon_{\text{Fermi}}(\text{X}) - E_0(\text{X})] \qquad (4)$$

where $\varepsilon_{\text{Fermi}}$ is the Fermi level, and $E_0$ is the vacuum level in the absence of surface polarity, as indicated in Fig. 2(b). The method to determine $E_0(x)$ in *ab initio* calculations can be seen in Supporting Information.

Thus, the core level difference between solids A and B in the absence of surface polarity is:

$$\Delta E_{\text{B-A}}(\text{CL}) = \Delta E_{\text{B'-A'}}(\text{CL}) + [\Delta E_{\text{A'-A}}(\text{CL}) - \Delta E_{\text{B'-B}}(\text{CL})] = \Delta E_{\text{B'-A'}}(\text{CL}) + \Delta E_{\text{B-A}}(\text{CL})_2, \qquad (5)$$

where $\Delta E_{\text{B-A}}(\text{CL})_2$ is the second correction for $\Delta E_{\text{B-A}}(\text{CL})$ due to lattice deformation.

By considering both Eq. 3 and Eq. 5, finally the core level difference between solids A and B involving the surface polarity and the lattice deformation can be calculated as:

$$\Delta E_{\text{B-A}}(\text{CL}) = \Delta E_{\text{B'-A'}}(\text{CL}) + \Delta E_{\text{B-A}}(\text{CL})_1 + \Delta E_{\text{B-A}}(\text{CL})_2. \qquad (6)$$

Then the calculation method for the band alignment between two solids based on XPS measurement procedure is proposed as:

$$\Delta E_{\text{B-A}}(\text{VBM}) = [\Delta E_{\text{VBM-CL}}(\text{B}) - \Delta E_{\text{VBM-CL}}(\text{A})] + \Delta E_{\text{B'-A'}}(\text{CL}) + \Delta E_{\text{B-A}}(\text{CL})_1 + \Delta E_{\text{B-A}}(\text{CL})_2. \qquad (7)$$

However, the last two terms were usually missed in the previous band-alignment calculations based on XPS measurement.

It is worth noting that valence band alignment using Eq. 7 depends on the distance $d$ between two solids in the junction supercell (see Fig. 1). Therefore two types of band alignment can be defined. When $d$ is large enough to cancel the interaction between two solids, their band alignment is intrinsic. While the two solids approach closer, the interaction between them become stronger, e.g. the charges can transfer between them to induce the dipole layer at their interface. In such a case, the band alignment is called the coupled.

To illustrate the reliability of Eq. 7, the band alignment between anatase and rutile phase $TiO_2$ will be taken as an example. $TiO_2$ is the mostly studied metal oxide as a prototypical model system for heterogeneous catalysis,[12-16] photochemistry,[17-19] surface science,[20-23] and so on.[24-26] On the other hand, the band alignment of two phases of $TiO_2$ is an active and controversial topic.[27-33] The XPS experiments observed that the band-edge positions of rutile $TiO_2$ are higher than those of anatase, whereas EIA showed the band edges of anatase straddle those of rutile. The calculation method Eq. 7 proposed by us can explain this experimental discrepancy, as demonstrated below.

For real $TiO_2$ crystals, the anatase (101) and rutile (110) surfaces account for the majorities of exposed surfaces respectively, which will be adopted for the band alignment. These two surfaces are modeled by the stoichiometric $p(3 \times 1)$ rutile $TiO_2(110)$ and $p(1 \times 2)$ anatase $TiO_2(101)$ periodically repeated slabs with the vacuum space of ~11 Å, respectively. All DFT+$U$ calculations are performed using Vienna *ab-initio* simulation package (VASP).[34, 35] The core level energies are calculated under the initial state approximation. The corresponding core state eigenenergies are taken from any one of unfixed Ti or O near the center of the slabs. The detailed settings of calculation can be found in Supporting Information.

The band alignments between anatase and rutile $TiO_2$ calculated by Eq. 7 are summarized in Table 1. For reference, some corresponding experimental observations are added into Table 1. In addition, the band alignments without the two corrections are also listed for comparison. Obviously, $\Delta E_{R-A}$(VBM) calculated without corrections exaggeratedly deviate from the experimental values, being no comparability with experimental values. Furthermore, it is physically unreasonable that the predicted values strongly depend on the selected core levels. In contrast, by taking the corrections implemented in Eq. 7, the predicted band offsets match the experimental ones very well. Besides, any selected core levels, e.g. Ti 1$s$, Ti 2$s$, Ti 2$p$, and O 1$s$, gives the identical values for band alignments in both the intrinsic and coupled conditions. This comparison implies that our proposed scenario for band alignment calculation (i.e. Eq. 7) is quite reliable.

The band alignment depends on the distance $d$ between two surface slabs may be seen in Table 1 for our model of $TiO_2$ junction. The detailed distance-dependent value of the band alignment is shown in Figure 3. When the two surface slabs are separated by a vacuum space of more than 5 Å (corresponding to the weak interaction between them), their VBM offset approaches a constant, called by us the intrinsic band alignment. When $d$ is less than 5 Å, the interaction between surface slabs is not negligible to the band offset. For example, when $d$ is 1.67 Å, which is half the width of a trilayer of anatase (see Figure S2 in Supporting Information), the band offset is significantly different from the intrinsic value at $d$ = 8.17 Å. The band alignment with the strong solid/solid interface interaction was called by us the coupled one.

Besides the XPS method, the EIA method can also characterize the band alignment between two solids, which corresponds to the aforementioned intrinsic type because EIA separately measures the VBM or CBM positions of semiconductors. The principle of EIA measurement is briefly summarized in Supporting Information. Our calculated intrinsic band alignment between two $TiO_2$ phases deviates somewhat from values of EIA measurements, which are listed in the last column of Table 1. This deviation may be caused by the surface adsorption of TiO$_2$ in the electrolyte solution and the compensation of surface polarity by adsorbates during the EIA measurements. Assuming the adsorbates on anatase (101) and rutile (110) surfaces completely compensate their surface polarity, the predicted $\Delta E_{R-A}$(VBM) should be -0.08 eV, very close to the EIA value of -0.04 eV.[36, 37]

**Table 1.** The calculated band alignments between rutile and anatase TiO$_2$ with/without the corrections. Different core levels are adopted as the energy reference, which should not affect the final results of band alignment physically. Some corresponding experimental values taken from literature are also presented for reference. All energy values are in unit of eV.

|  |  | Ti 1s | Ti 2s | Ti 2p | O 1s | XPS | EIA* |
|---|---|---|---|---|---|---|---|
| $\Delta E_{VBM-CL}$(R) − $\Delta E_{VBM-CL}$(A) |  | -0.98 | -0.94 | -0.95 | -0.98 |  |  |
| $\Delta E_{R'-A'}$(CL) | a | -3.81 | -3.41 | -3.46 | -1.37 |  |  |
|  | b | -3.54 | -3.14 | -3.19 | -1.09 |  |  |
| $\Delta E_{R-A}$(CL)$_1$ |  |  | 0.26 |  |  |  |  |
| $\Delta E_{R-A}$(CL)$_2$ |  | 4.71 | 4.28 | 4.33 | 2.27 |  |  |
| $\Delta E_{R-A}$(VBM) | a' | -4.79 | -4.35 | -4.41 | -2.35 | 0.39[30] |  |
|  | a | 0.18 | 0.19 | 0.18 | 0.18 | 0.7[38] | -0.04[A(101)[36]; R(110)[37]] |
|  | b' | -4.52 | -4.08 | -4.14 | -2.07 | 0.37[39] |  |
|  | b | 0.45 | 0.46 | 0.45 | 0.46 |  |  |

* The values in this column are calculated using Eq. S5 in Supporting Information.

a –values calculated using Eq. 7 when the distance d between two slabs R and A is 8.17 Å.

b –values calculated using Eq. 7 at d = 1.67 Å.

a′ –values calculated without corrections at d = 8.17 Å.

b′ –values calculated without corrections at d = 1.67 Å.

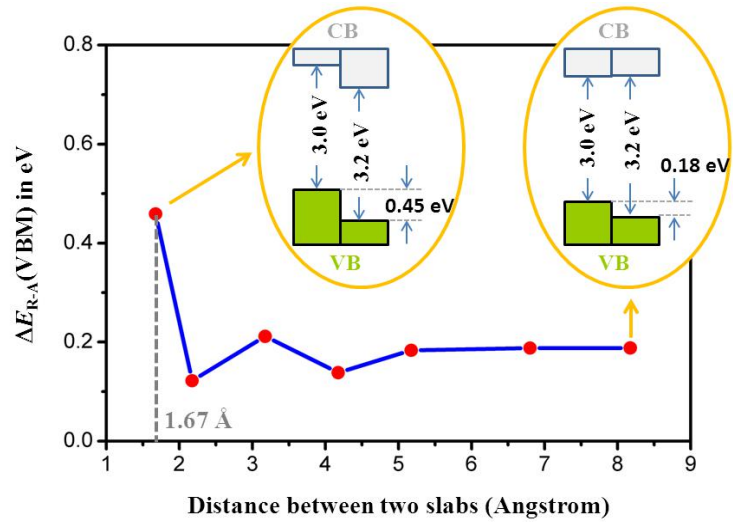

**Figure 3.** The band alignment between anatase and rutile TiO$_2$ varies as the distance $d$ between two slabs.

In contrast, the XPS measurement of the band alignment between two solids involves the solid/solid interface effect and thus corresponds to the coupled band alignment. Since the coupled band alignment is sensitive to the interfacial dipole layer which may be created by a change in the interfacial ions' configuration, charge transfer between two solids, etc., it is reasonable that the XPS experimental results of the band alignment between anatase and rutile TiO$_2$ are a little scattered. In fact, previous studies of the strain effect on the band structure of semiconductors had noticed that the change in interfacial atomic configuration, i.e. change of bond lengths, bond angles between interfacial atoms, does not lead to a much big shift in VBMs of solids with an indeterminate sign:[40] the prolonged bond lengths decrease the bonding-antibonding splitting and the bandwidth, respectively increasing and decreasing the VBM energy, whereas, the shortened bond lengths behave inversely. In this sense, the band alignment calculated here without performing structural relaxation of the A'/B' junction model to meet the elusive real interfacial atomic picture is a good approximation. Taking half the atomic periodicity of the longer-periodicity solid in the heterojunction along the $z$ direction as the minimum value of $d$ is reasonable to make overlapping of the atomic orbitals correct. Half the width of a trilayer of anatase (rutile) is 1.67 (1.59) Å (see Figure S2 in Supporting Information). Our calculated VBM band offset between anatase and rutile at $d$ = 1.67 Å is 0.45 eV, which locates in the range of XPS values as expected (see Table 1). So, our coupled band alignment at the minimum value of $d$ may be acted as a better approximation than the first approximation (the intrinsic band alignment) that commonly is used to help design of devices with solid interfaces.

In summary, an *ab-initio* method for reliable calculation of band alignment between two solids has been proposed on the basis of the protocol of XPS measurement with two corrections in the core level from surface polarity and lattice deformation. Our method can yield two types of band alignment. The intrinsic band alignment free of the interfacial interaction between two solids corresponds to the band alignments measured by the EIA method, while the coupled one containing the interfacial interaction corresponds to the XPS measurement. The intrinsic and coupled band alignments between rutile and anatase $TiO_2$ calculated by our method are different but in well agreement with corresponding experimental observations. The theoretical principle of our work is independent of the type of material, and it may predict the band alignments between any two solids.

## ASSOCIATED CONTENT

**Supporting Information**

This material is available free of charge via the Internet at http://pubs.acs.org.

The method to determine $E_0$; Methods and surface models; The band edge positions determined by EIA.

## AUTHOR INFORMATION

**Corresponding Author**

*E-mail: zhangdaoyu@seu.edu.cn

*E-mail: sdong@seu.edu.cn. Tel. & Fax: +86 25 52090606.

**Notes**

The authors declare no competing financial interest.

## ACKNOWLEDGMENT

Work was supported by the National Natural Science Foundation of China (Grant No. 11674055).

## REFERENCE

(1) Peressi, M.; Binggeli, N.; Baldereschi, A. Band Engineering at Interfaces: Theory and Numerical Experiments. *J. Phys. D Appl. Phys.* **1998**, *31*, 1273-1299.

(2) Van de Walle, C. G.; Neugebauer, J. Universal Alignment of Hydrogen Levels in Semiconductors, Insulators and Solutions. *Nature* **2003**, *423*, 626-628.

(3) Walsh, A.; Butler, K. T. Prediction of Electron Energies in Metal Oxides. *Account. Chem. Res.* **2014**, *47*, 364-372.

(4) Wei, S. H.; Zunger, A. Role of d-Orbitals in Valence-Band Offsets of Common-Anion Semiconductors. *Phys. Rev. Lett.* **1987**, *59*, 144-147.

(5) Zhang, Z.; Yates, J. T. Band Bending in Semiconductors: Chemical and Physical Consequences at Surfaces and Interfaces. *Chem. Rev.* **2012**, *112*, 5520-5551.

(6) Kraut, E. A.; Grant, R. W.; Waldrop, J. R.; Kowalczyk, S. P. Precise Determination of the Valence-Band Edge in X-Ray Photoemission Spectra - Application to Measurement of Semiconductor Interface Potentials. *Phys. Rev. Lett.* **1980**, *44*, 1620-1623.

(7) Wei, S. H.; Zunger, A. Predicted Band-Gap Pressure Coefficients of All Diamond and Zinc-Blende Semiconductors: Chemical Trends. *Phys. Rev. B* **1999**, *60*, 5404-5411.

(8)	Chambers, S. A.; Ohsawa, T.; Wang, C. M.; Lyubinetsky, I.; Jaffe, J. E. Band Offsets at the Epitaxial Anatase $TiO_2$/N-$SrTiO_3$(001) Interface. *Surf. Sci.* **2009**, *603*, 771-780.

(9)	Gai, Y. Q.; Li, J. B.; Li, S. S.; Xia, J. B.; Wei, S. H. Design of Narrow-Gap TiO(2): A Passivated Codoping Approach for Enhanced Photoelectrochemical Activity. *Phys. Rev. Lett.* **2009**, *102*, 036402.

(10)	Ju, M.-G.; Sun, G.; Wang, J.; Meng, Q.; Liang, W. Origin of High Photocatalytic Properties in the Mixed-Phase $TiO_2$: A First-Principles Theoretical Study. *ACS Appl. Mater. Inter.* **2014**, *6*, 12885-12892.

(11)	Ma, X.; Dai, Y.; Yu, L.; Huang, B. Interface Schottky Barrier Engineering Via Strain in Metal-Semiconductor Composites. *Nanoscale* **2016**, *8*, 1352-1359.

(12)	Babu, V. J.; Vempati, S.; Uyar, T.; Ramakrishna, S. Review of One-Dimensional and Two-Dimensional Nanostructured Materials for Hydrogen Generation. *Phys. Chem. Chem. Phys.* **2015**, *17*, 2960-2986.

(13)	Chen, X.; Liu, L.; Huang, F. Black Titanium Dioxide ($TiO_2$) Nanomaterials. *Chem. Soc. Rev.* **2015**, *44*, 1861-1885.

(14)	Li, J.; Wu, N. Semiconductor-Based Photocatalysts and Photoelectrochemical Cells for Solar Fuel Generation: A Review. *Catal. Sci. Technol.* **2015**, *5*, 1360-1384.

(15)	Liu, B.; Zhao, X.; Terashima, C.; Fujishima, A.; Nakata, K. Thermodynamic and Kinetic Analysis of Heterogeneous Photocatalysis for Semiconductor Systems. *Phys. Chem. Chem. Phys.* **2014**, *16*, 8751-8760.

(16)	Moniz, S. J. A.; Shevlin, S. A.; Martin, D. J.; Guo, Z.-X.; Tang, J. Visible-Light Driven Heterojunction Photocatalysts for Water Splitting - a Critical Review. *Energ. Environ. Sci.* **2015**, *8*, 731-759.

(17)	Kalyanasundaram, K., Photochemical Applications of Solar Energy: Photocatalysis and Photodecomposition of Water. In *Photochemistry: Volume 41*, The Royal Society of Chemistry: 2013; Vol. 41, pp 182-265.

(18)	Wu, Z.; Zhang, W.; Xiong, F.; Yuan, Q.; Jin, Y.; Yang, J.; Huang, W. Active Hydrogen Species on $TiO_2$ for Photocatalytic $H_2$ Production. *Phys. Chem. Chem. Phys.* **2014**, *16*, 7051-7057.

(19)	Zehr, R. T.; Henderson, M. A. Thermal Chemistry and Photochemistry of Hexafluoroacetone on Rutile $TiO_2$(110). *Phys. Chem. Chem. Phys.* **2010**, *12*, 8085-8092.

(20)	Agosta, L.; Zollo, G.; Arcangeli, C.; Buonocore, F.; Gala, F.; Celino, M. Water Driven Adsorption of Amino Acids on the (101) Anatase $TiO_2$ Surface: An Ab Initio Study. *Phys. Chem. Chem. Phys.* **2015**, *17*, 1556-1561.

(21)	Guo, Q.; Zhou, C.; Ma, Z.; Ren, Z.; Fan, H.; Yang, X. Elementary Photocatalytic Chemistry on $TiO_2$ Surfaces. *Chem. Soc. Rev.* **2016**, *45*, 3701-3730.

(22)	Lun Pang, C.; Lindsay, R.; Thornton, G. Chemical Reactions on Rutile $TiO_2$(110). *Chem. Soc. Rev.* **2008**, *37*, 2328-2353.

(23)	Zhang, D. Y.; Yang, M. N.; Dong, S. Improving the Photocatalytic Activity of $TiO_2$ through Reduction. *RSC Adv.* **2015**, *5*, 35661-35666.

(24)	Li, H. F.; Yu, H. T.; Quan, X.; Chen, S.; Zhao, H. M. Improved Photocatalytic Performance of Heterojunction by Controlling the Contact Facet: High Electron Transfer Capacity between $TiO_2$ and the {110} Facet of $BiVO_4$ Caused by Suitable Energy Band Alignment. *Adv. Funct. Mater.* **2015**, *25*, 3074-3080.

(25)	Zhang, D. Y.; Yang, M. N. Band Structure Engineering of $TiO_2$ Nanowires by N-P Codoping for Enhanced Visible-Light Photoelectrochemical Water-Splitting. *Phys. Chem. Chem. Phys.* **2013**, *15*, 18523-18529.

(26)	Zhang, Y.; Jiang, Z.; Huang, J.; Lim, L. Y.; Li, W.; Deng, J.; Gong, D.; Tang, Y.; Lai, Y.; Chen, Z. Titanate and Titania Nanostructured Materials for Environmental and Energy Applications: A Review. *RSC Adv.* **2015**, *5*, 79479-79510.

(27)	Kullgren, J.; Aradi, B.; Frauenheim, T.; Kavan, L.; Deak, P. Resolving the Controversy About the

Band Alignment between Rutile and Anatase: The Role of OH$^-$/H$^+$ Adsorption. *J. Phys. Chem. C* **2015**, *119*, 21952-21958.

(28) Mi, Y.; Weng, Y. Band Alignment and Controllable Electron Migration between Rutile and Anatase TiO$_2$. *Sci. Rep.* **2015**, *5*, 11482-11482.

(29) Nosaka, Y.; Nosaka, A. Y. Reconsideration of Intrinsic Band Alignments within Anatase and Rutile TiO$_2$. *J. Phys. Chem. Lett.* **2016**, *7*, 431-434.

(30) Scanlon, D. O.; Dunnill, C. W.; Buckeridge, J.; Shevlin, S. A.; Logsdail, A. J.; Woodley, S. M.; Catlow, C. R. A.; Powell, M. J.; Palgrave, R. G.; Parkin, I. P., et al. Band Alignment of Rutile and Anatase TiO$_2$. *Nat. Meter.* **2013**, *12*, 798-801.

(31) Zhang, D. Y.; Yang, M. N.; Dong, S. Electric-Dipole Effect of Defects on the Energy Band Alignment of Rutile and Anatase TiO$_2$. *Phys. Chem. Chem. Phys.* **2015**, *17*, 29079-29084.

(32) Deak, P.; Aradi, B.; Frauenheim, T. Band Lineup and Charge Carrier Separation in Mixed Rutile-Anatase Systems. *J. Phys. Chem. C* **2011**, *115*, 3443-3446.

(33) Wang, C.; Zhang, X.; Wei, Y.; Kong, L.; Chang, F.; Zheng, H.; Wu, L.; Zhi, J.; Liu, Y. Correlation between Band Alignment and Enhanced Photocatalysis: A Case Study with Anatase/ TiO$_2$(B) Nanotube Heterojunction. *Dalton Trans.* **2015**, *44*, 13331-13339.

(34) Blochl, P. E. Projector Augmented-Wave Method. *Phys. Rev. B* **1994**, *50*, 17953-17979.

(35) Kresse, G.; Furthmuller, J. Efficient Iterative Schemes for Ab Initio Total-Energy Calculations Using a Plane-Wave Basis Set. *Phys. Rev. B* **1996**, *54*, 11169-11186.

(36) Kavan, L.; Gratzel, M.; Gilbert, S. E.; Klemenz, C.; Scheel, H. J. Electrochemical and Photoelectrochemical Investigation of Single-Crystal Anatase. *J. Am. Chem. Soc.* **1996**, *118*, 6716-6723.

(37) Nakamura, R.; Ohashi, N.; Imanishi, A.; Osawa, T.; Matsumoto, Y.; Koinuma, H.; Nakato, Y. Crystal-Face Dependences of Surface Band Edges and Hole Reactivity, Revealed by Preparation of Essentially Atomically Smooth and Stable (110) and (100) N-TiO$_2$ (Rutile) Surfaces. *J. Phys. Chem. B* **2005**, *109*, 1648-1651.

(38) Pfeifer, V.; Erhart, P.; Li, S.; Rachut, K.; Morasch, J.; Broetz, J.; Reckers, P.; Mayer, T.; Ruehle, S.; Zaban, A., et al. Energy Band Alignment between Anatase and Rutile TiO$_2$. *J. Phys. Chem. Lett.* **2013**, *4*, 4182-4187.

(39) Xiong, G.; Shao, R.; Droubay, T. C.; Joly, A. G.; Beck, K. M.; Chambers, S. A.; Hess, W. P. Photoemission Electron Microscopy of TiO$_2$ Anatase Films Embedded with Rutile Nanocrystals. *Adv. Funct. Mater.* **2007**, *17*, 2133-2138.

(40) Li, Y. H.; Gong, X. G.; Wei, S. H. Ab Initio Calculation of Hydrostatic Absolute Deformation Potential of Semiconductors. *Appl. Phys. Lett.* **2006**, *88*, 042104.

**Supporting Information for**

**Translating XPS Measurement Procedure for Band Alignment into Reliable *Ab Initio* Calculation Method**

Daoyu Zhang,*,† Minnan Yang,‡ Huimin Gao,† and Shuai Dong*,†

†School of Physics, Southeast University, Nanjing, 211189, China.

‡Department of Physics, China Pharmaceutical University, Nanjing, 211198, China.

**SI-1.** The method to determine $E_0$

The solid with a surface was modeled by a slab with a vacuum space. When the average electrostatic potential over the whole model lattice is taken as the zero energy reference, the electric dipole layer representing the surface polarity creates the electrostatic potential distribution as sketched in Figure S1(a). The electrostatic potential in the right side of the surface (i.e. vacuum region) increases by $E^{VR}$ and it decreases by $E^{SR}$ in the left side (i.e. solid region). Usually the polarized charges are localized at the outermost surficial ions,[1-3] so the width of the equivalent capacitor is quite narrow, which can be neglected comparing with the width of the vacuum space and the slab. Thus, one can obtain:

$$E^{SR} \times w_{\text{solid}} = E^{VR} \times w_{\text{vacuum}} , \qquad (S1)$$

where $w_{\text{solid}}$ and $w_{\text{vacuum}}$ denote widths of the slab and the vacuum respectively.

The solid line shown in Figure S1(b) is a typical DFT-calculated electrostatic potential of a slab lattice with surface polarity, which has been averaged in the *x-y* plane parallel to the surface. $E_R$ and $E_L$ is the maximum and minimum values of the electrostatic potential in the vacuum region, and their difference is equal to $\varphi$, the electrostatic potential difference between the solid and the vacuum induced by the surface polarity. The dashed line is the electrostatic potential of the slab in the absence of surface polarity. $E_0$, the vacuum level of the slab lattice in the absence of surface polarity, can be calculated by:

$$E_0 = E_L + \varphi \times w_{\text{vacuum}}/(w_{\text{solid}} + w_{\text{vacuum}}). \qquad (S2)$$

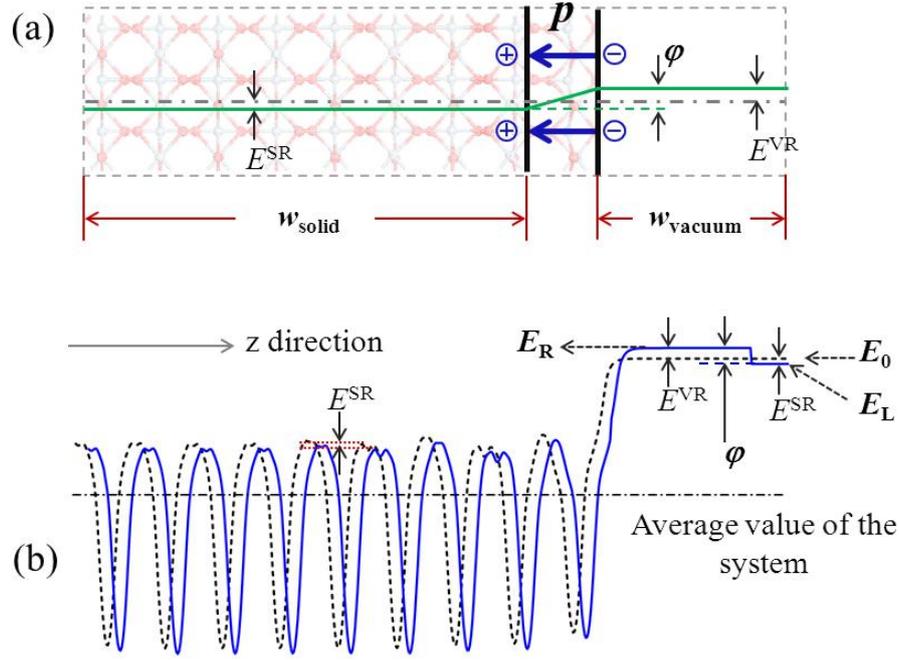

**Figure S1.** (a) The electrostatic potential created by the surface dipole layer, when the average electrostatic potential over the whole slab lattice is taken as the energy reference. (b) Solid line: the potential calculated by DFT for the slab lattice. Dashed line: the one deducting (a) from the solid line. The potentials have been averaged in the *x-y* plane parallel to the surface.

**SI-2.** Methods and surface models

All DFT+$U$ calculations are performed using the projector-augmented wave pseudopotentials as implemented in the Vienna *ab initio* Simulation Package (VASP)[4, 5] within the local-density approximation (LDA). The energy cutoff for plane wave basis sets is 450 eV. The atomic positions and cell parameters are relaxed until the forces on each atom are less than 0.01 eV/Å, and the self-consistent convergence accuracy is set at $1\times10^{-5}$ eV/atom. The experimental lattice constants, $a=b=3.785$ Å, $c=9.502$ Å for primitive anatase phase $TiO_2$, and $a=4.593$ Å, $c=2.959$ Å for rutile phase $TiO_2$,[6] are used to build the surface slab models. For integration within the first Brillouin zone, $\Gamma$-point sampling was selected. On-site Coulomb repulsion $U=3.5$ eV is imposed on the $3d$ orbitals of Ti.

To model the surfaces, the stoichiometric $p(3\times1)$ rutile $TiO_2$ (110) plane and $p(1\times2)$ anatase $TiO_2$ (101) plane are adopted, respectively. Periodically repeated slabs with a vacuum space of ~11 Å. The number of atoms in the rutile (anatase) slab is 162 (192). Atoms of four trilayers in the anatase and rutile surface models were fixed at their bulk positions, as shown in Figure S2. The monopole, dipole, and quadrupole corrections are applied to the electrostatic interaction between the slab and its periodic images in the direction perpendicular to the slab.

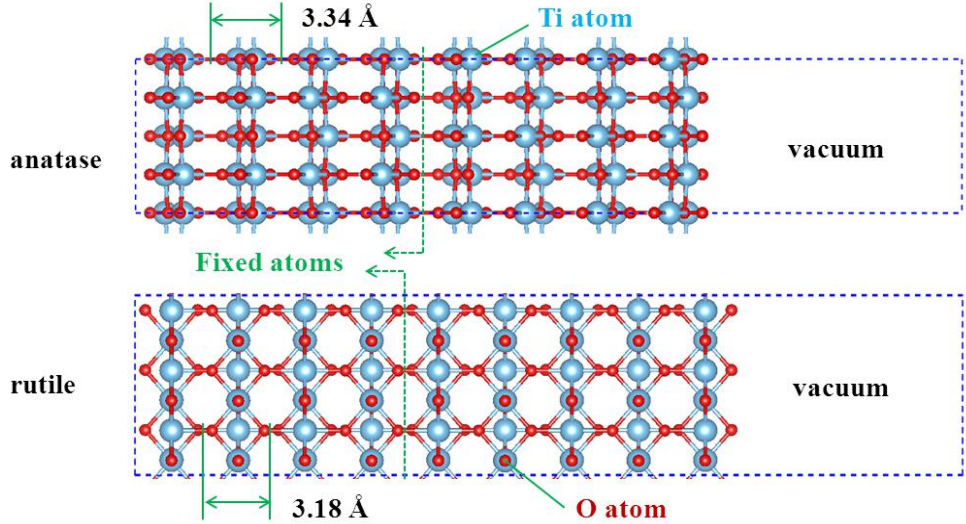

**Figure S2.** The side view of the anatase TiO$_2$ (101) and rutile TiO$_2$ (110) surface models. The width of the trilayer of anatase and rutile are 3.34 and 3.18 Å, respectively.

**SI-3.** The band edge positions determined by EIA

EIA measurements can determine the flat band potential ($U_{ft}$) of a semiconductor and the electrode potential relative to a reference electrode (e.g. the normal hydrogen electrode, NHE) when the space charge layer of the semiconductor becomes zero. $U_{ft}$ links the energy levels of the semiconductor and electrolyte by the relationship:[7, 8]

$$U_{ft} = \chi + \Delta E_F + V_H + \Phi_0, \tag{S3}$$

where $\chi$ is the electron affinity of the semiconductor; $\Delta E_F$ is the difference between the Fermi level and majority carrier band edge of the semiconductor; $V_H$ is the potential drop across the Helmholtz layer; and $\Phi_0$ is the scale factor (e.g. 4.5 eV for NHE). When $V_H$ equals zero at zero point of charge (pH$_{ZPC}$), the flat band potential ($U_{ft}^0$) is the intrinsic Fermi level of the semiconductor.

For metal oxides, the flat band potential varies with the pH value of solution following a linear relation known as the Nernstian relation:

$$U_{ft}^{pH} = U_{ft}^0 + 0.059(pH_{ZPC} - pH). \tag{S4}$$

Using measured data of $U_{fb}^{pH}$ and $E_g$ ($E_g$ is the band gap), the valence band edge energy of an n-type semiconductor with respect to the reference electrode can be calculated as:

$$E_{VBE} = U_{fb}^{pH} - \Delta E_F + E_g - 0.059(pH_{ZPC} - pH). \quad (S5)$$

To determine the value of $\Delta E_F$, experimentally certain impurities are introduced into the semiconductor, which can shift the Fermi level so close to the majority carrier band edge that $\Delta E_F$ is approximately zero and may be neglected in Eq. (S5). The valence band edges in two n-type semiconductors aligned by Eq. (S5) are apparently the intrinsic type.

The flat band potential of anatase $TiO_2$ (101) surface relative to the saturated calomel electrode (SCE) is -0.4 eV at pH = 0,[9] while that of rutile $TiO_2$ (110) surface is -0.25 eV at pH = 1.[10] Using Eq. (S5) and considering the band gaps (3.2 eV for anatase and 3.0 eV for rutile), the valence band edge energy of rutile can be estimated, which is lower by 0.04 eV than that of anatase.

## References

(1) Chretien, S.; Metiu, H. Electronic Structure of Partially Reduced Rutile Tio2(110) Surface: Where Are the Unpaired Electrons Located? *J. Phys. Chem. C* **2011**, *115*, 4696-4705.

(2) Deskins, N. A.; Rousseau, R.; Dupuis, M. Localized Electronic States from Surface Hydroxyls and Polarons in Tio2(110). *J. Phys. Chem. C* **2009**, *113*, 14583-14586.

(3) Zhang, D. Y.; Yang, M. N.; Dong, S. Hydroxylation of the Rutile Tio2(110) Surface Enhancing Its Reducing Power for Photocatalysis. *J. Phys. Chem. C* **2015**, *119*, 1451-1456.

(4) Blochl, P. E. Projector Augmented-Wave Method. *Phys. Rev. B* **1994**, *50*, 17953-17979.

(5) Kresse, G.; Furthmuller, J. Efficient Iterative Schemes for Ab Initio Total-Energy Calculations Using a Plane-Wave Basis Set. *Phys. Rev. B* **1996**, *54*, 11169-11186.

(6) Burdett, J. K.; Hughbanks, T.; Miller, G. J.; Richardson, J. W.; Smith, J. V. Structural Electronic Relationships in Inorganic Solids - Powder Neutron-Diffraction Studies of the Rutile and Anatase Polymorphs of Titanium-Dioxide at 15 and 295-K. *J. Am. Chem. Soc.* **1987**, *109*, 3639-3646.

(7) Nozik, A. J. Photoelectrochemistry - Applications to Solar-Energy Conversion. *Annu. Rev. Phys. Chem.* **1978**, *29*, 189-222.

(8) Xu, Y.; Schoonen, M. A. A. The Absolute Energy Positions of Conduction and Valence Bands of

Selected Semiconducting Minerals. *Am. Mineral.* **2000**, *85*, 543-556.

(9) Kavan, L.; Gratzel, M.; Gilbert, S. E.; Klemenz, C.; Scheel, H. J. Electrochemical and Photoelectrochemical Investigation of Single-Crystal Anatase. *J. Am. Chem. Soc.* **1996**, *118*, 6716-6723.

(10) Nakamura, R.; Ohashi, N.; Imanishi, A.; Osawa, T.; Matsumoto, Y.; Koinuma, H.; Nakato, Y. Crystal-Face Dependences of Surface Band Edges and Hole Reactivity, Revealed by Preparation of Essentially Atomically Smooth and Stable (110) and (100) N-Tio2 (Rutile) Surfaces. *J. Phys. Chem. B* **2005**, *109*, 1648-1651.